\documentclass[superscriptaddress, preprintnumbers,noeprint,aps,amsmath,amssymb,prl]{revtex4-2}

\usepackage[utf8]{inputenc}

\usepackage{graphicx}
\usepackage{xspace}
\usepackage{soul}
\usepackage{wrapfig}

\newcommand{\calF}{{\cal F}}

\newcommand{\ie}{{\it i.e.}\xspace}

\newcommand{\eg}{{\it e.g.}\xspace}
\newcommand{\GeV}{\text{GeV}}

\usepackage{color}
\usepackage[makeroom]{cancel}
\usepackage[normalem]{ulem}
\newcommand{\Phadr}{P_h}
\newcommand{\Pelec}{P_e}
\definecolor{pkcolor}{rgb}{0,0.1,0.7}

\newcommand\pkout{\marginpar{\color{pkcolor}$\clubsuit$}\bgroup\markoverwith{\color{pkcolor}{\rule[2pt]{2pt}{0.8pt}}}\ULon}

\definecolor{sscolor}{rgb}{0.0,0.5,0.5}

\newcommand\ssout{\marginpar{\color{sscolor}$\otimes$}\bgroup\markoverwith{\color{sscolor}{\rule[2pt]{2pt}{0.8pt}}}\ULon}

\definecolor{kkcolor}{rgb}{0.7,0.1,0.}

\newcommand\kkout{\marginpar{\color{kkcolor}$\int$}\bgroup\markoverwith{\color{kkcolor}{\rule[04ex]{2pt}{0.8pt}}}\ULon}

\begin{document}

\preprint{IFJPAN-IV-2021-10}

\title{Probing gluon number density with electron-dijet correlations at EIC}

\author{Andreas van Hameren}
\affiliation{Institute of Nuclear Physics, Polish Academy of Sciences,
             Radzikowskiego 152, 31-342 Krak\'ow, Poland}

\author{Piotr Kotko}
\affiliation{AGH University of Science and Technology, Physics Faculty,
             Mickiewicza 30, 30-059 Krak\'ow, Poland}

\author{Krzysztof Kutak}
\affiliation{Institute of Nuclear Physics, Polish Academy of Sciences,
             Radzikowskiego 152, 31-342 Krak\'ow, Poland}

\author{Sebastian Sapeta}
\affiliation{Institute of Nuclear Physics, Polish Academy of Sciences,
             Radzikowskiego 152, 31-342 Krak\'ow, Poland}

\author{Elżbieta Żarów}
\affiliation{AGH University of Science and Technology, Physics Faculty,
             Mickiewicza 30, 30-059 Krak\'ow, Poland}

\begin{abstract}
We propose a novel way of studying the gluon number density (the so-called
Weizs\"acker-Williams  gluon distribution) using the planned Electron Ion Collider. Namely, with the help of the azimuthal correlations between 
the total transverse momentum of the dijet system and the scattered electron, we
examine an interplay between the effect of the soft gluon emissions (the Sudakov form factor) and the gluon saturation effects.
The kinematic cuts are chosen such that the dijet system is produced in the
forward direction in the laboratory frame, which provides an upper
bound on the probed longitudinal fractions of the hadron momentum carried by
scattered gluons. Further cuts enable us to use the factorization formalism that
directly involves the unpolarized Weizs\"acker-Williams gluon distribution. We find this  observable to be very sensitive to the soft gluon emission and moderately sensitive to the gluon saturation.
 
\end{abstract}

\maketitle

High energy \emph{deep inelastic scattering}~(DIS) of electrons and
nuclei at the future Electron Ion Collider~(EIC)~\cite{Accardi:2012qut} will
provide a unique opportunity to perform detailed studies of various aspects of
Quantum Chromodynamics~(QCD)~\cite{AbdulKhalek:2021gbh}. Amongst them is a
formation of nuclei in terms of QCD degrees of freedom and, in particular, the
dynamics of strongly correlated gluon systems, which leads to emergence of the
phenomenon of saturation \cite{Gribov:1984tu,Mueller:1985wy}.

An especially interesting part of the EIC program concerns jet production
\cite{Arratia:2019vju}. It  will allow to study in more detail the internal
structure of various hadronic targets, such as $p$, $Pb$, $Au$
\cite{Liu:2018trl,delCastillo:2020omr}, jet shapes, transport properties of cold
nuclear matter \cite{Li:2020rqj} low $x$ effects including gluon saturation \cite{Kolbe:2020tlq},
effects which are not all accessible in inclusive processes. As it has been recognized
already in HERA-related studies
\cite{Askew:1994is,Bartels:1996gr,Bartels:1996wx,Kwiecinski:1997ga,Kwiecinski:1999wj,Aaron:2011ef}
(see also \cite{AbelleiraFernandez:2012cc}), the most suitable processes for the
low $x$ studies are those in which jets are produced in the forward direction
with respect to the incoming electron, that is they have large rapidities.
Kinematically, this allows for a better focus on the events
in which
the partons extracted from the proton or nucleus carry longitudinal momentum fraction,
$x$, which is small enough for the saturation physics to be applicable and
tested with high precision.

In the present, work we focus on the production of at least two jets in DIS
collisions. Such processes
are especially interesting, because, unlike inclusive processes, at
high energies, they are directly sensitive to the gluon number density in
hadrons, the so-called Weizs\"acker-Williams transverse momentum dependent (TMD)
gluon distribution~\cite{Kharzeev:2003wz,Dominguez:2011wm}.

The forward dijet production setup can be used reliably only if our
theoretical predictions for that process reach certain quality. Specifically,
the above kinematic configuration leads to appearance of two groups of 
potentially
large logarithms: $\ln x$ and $\ln \mu $, where the latter, with $\mu$ being of the order
of the transverse momenta of the jets, are called the Sudakov logarithms.
 Both types of logarithms should be resummed
 simultaneously and such resummation can be performed, as shown in
 Refs.~\cite{Mueller:2013wwa, Mueller:2012uf},  where the results were obtained
 in position space, using the color dipole formalism \cite{Mueller:1993rr}.

In this work we apply the recently formulated small-$x$ Improved Transverse
Momentum Dependent (ITMD) factorization \cite{Kotko:2015ura,vanHameren:2016ftb}
which generalizes the small-$x$ TMD factorization \cite{Dominguez:2011wm} to
account for  power corrections, so that it reduces to  the ordinary $k_T$
factorization \cite{Catani:1990eg} when the saturation effects are neglected.
This framework, together with a model for the Sudakov resummation, has recently
been successfully applied to shape description of dijet azimuthal angle
decorrelations data in $p-p$ and $p-Pb$ collisions \cite{vanHameren:2019ysa}.  ITMD
can be regarded as a special case of the Color Glass
Condensate~(CGC)~\cite{Gelis:2010nm} description once one neglects multiple
partonic interactions (the so-called genuine twists) \cite{Altinoluk:2019fui}.
For a detailed study of dijet production at EIC within the full CGC framework see \cite{Mantysaari:2019hkq}.

It is worth mentioning, that higher genuine twits constitute only one of the mechanisms of
saturation \cite{Altinoluk:2019wyu}. The difference between higher genuine twists
and  the two-body contribution is rather subtle
\cite{Fujii:2020bkl,BoussarieEIC2021} and is accessible in rather small
transverse momentum domain. It follows, that the higher genuine twists are suppressed
at large transverse momenta 
or photon virtualities, $Q^2$. Therefore the
ITMD framework is ideally suited for inclusive dijets studies, provided suitable
kinematic cuts are imposed.  A further advantage of this formalism is that it
operates directly in momentum space and therefore allows to relax kinematic
approximations often used in the coordinate space calculations.  This is
especially important for Monte Carlo implementations, which, in particular,
allow for a very natural and flexible application of cuts. 

In this Letter we propose 
to utilize the azimuthal correlations between the forward dijet system and the scattered electron  at EIC as a tool to study the gluon number distribution. This observable has not been studied in detail
in the context of the saturation physics so far.
Since there are three tagged final states -- the scattered electron and
two jets -- there is a variety of potentially interesting observables. In studies
of the TMD gluon distributions, especially in the saturation regime, the most
interesting observables concern azimuthal correlations between the collision
products. A quite standard choice is to look at the azimuthal angle between the
jets
\cite{Marquet:2007vb,Deak:2009xt,Deak:2010gk,Kutak:2012rf,vanHameren:2016ftb} or
hadrons~\cite{Albacete:2010pg,Zheng:2014vka}. It is obviously the only choice
for the dijets at LHC, whereas for the DIS processes, one typically decouples
the photon flux from the electron, and thus such observable seems also quite
natural. However, in the approaches involving the TMD parton distributions, the
transverse momenta of gluons extracted from the proton or nuclei (both internal
and due to Sudakov-like soft emissions)  are balanced by the whole final state, including the electron.
As
will be demonstrated in the following sections, the angle between the dijet system and the electron   is very sensitive
to soft gluon emissions
and visibly sensitive to saturation effects
Hence, it offers a particularly good handle on
quantifying the role Sudakov resummation in dijet production in
DIS and its interplay with saturation effects.
In general, observables involving scattered electron did not draw a lot of attention in the small-$x$ community so far. Various asymmetries related to the EIC physics have been discussed on general ground in \cite{Boer:2016fqd}. Quite recently, in \cite{Mantysaari:2020lhf} the Authors considered azimuthal electron-vector meson (or photon) correlations in exclusive diffractive production within CGC framework.

\section{Framework}

\begin{figure}[t]
  \begin{center}
    \includegraphics[width=0.33\textwidth]{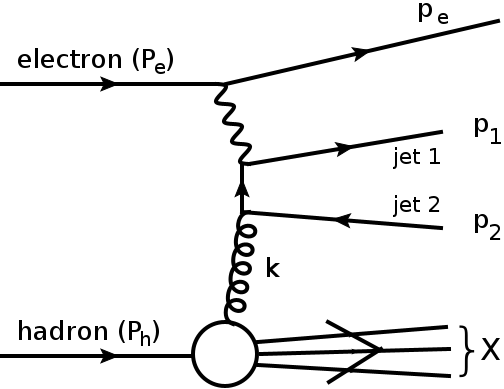}
    \hspace{13ex}
    \includegraphics[width=0.38\textwidth]{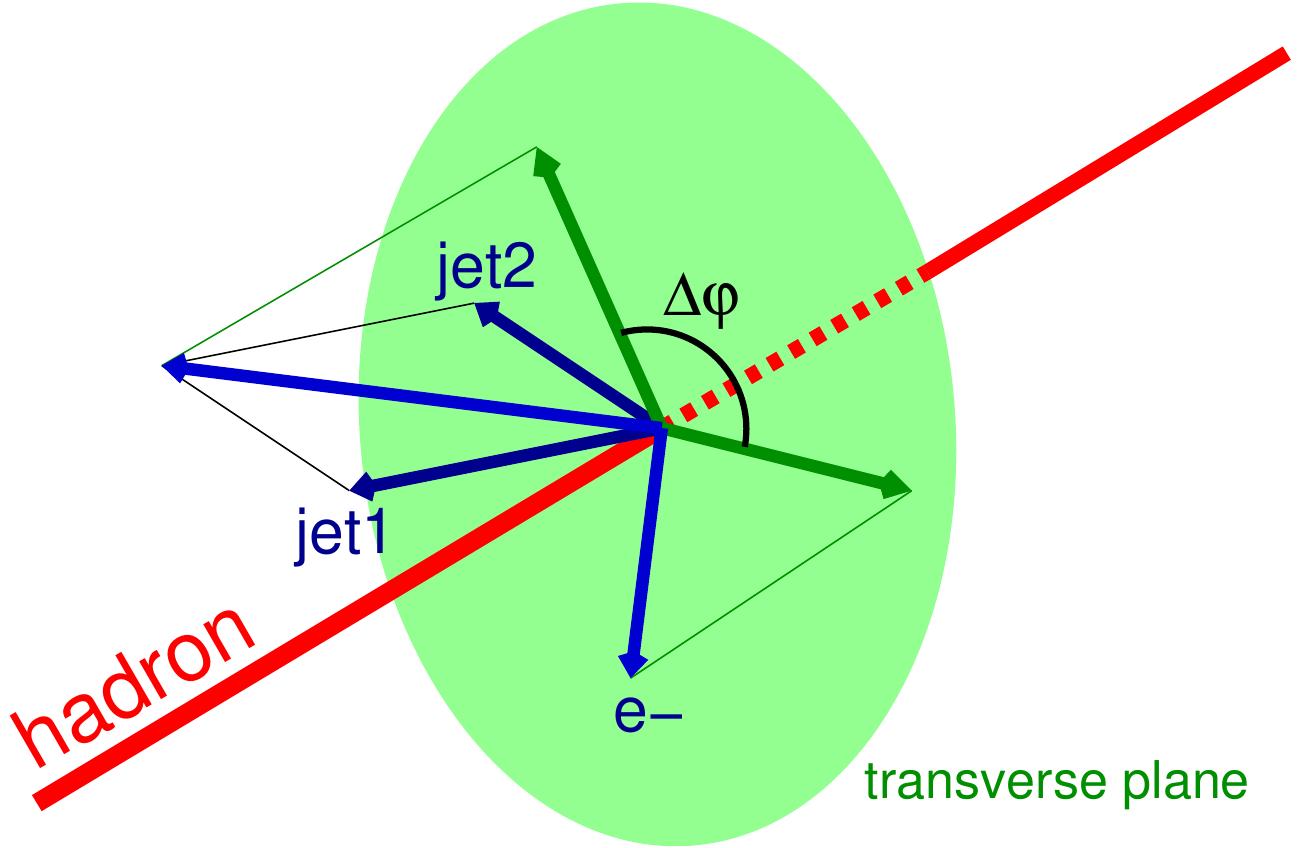}
  \end{center}
  \caption{Dijet production in DIS collision (left), and a graphical representation of the studied observable $\Delta\varphi(J_1+J_2,e^-)$ (right).
  }
  \label{fig:diag}
\end{figure}

We consider the process of production of a dijet system in DIS
\begin{equation}
  e+h\rightarrow e'+J_1+J_2+X\,,
  \label{eq:proc-def}
\end{equation}
where $h$ can be a proton or an ion. The corresponding diagram is depicted on the left of
Fig.~\ref{fig:diag}, where we also define momenta of the particles involved in
the process.


In our calculations, we shall use the ITMD framework, which, as explained in the
Introduction, is suitable for our process of interest.
It accounts for a complete kinematical twist, which amounts to resumming the
$(Q_s/k_T^2)^n$ and $(k_T^2/P^2_T)^n$ contributions, where $P_T \sim p_{T1},
p_{T2}$, which has been shown by a derivation within 
CGC~\cite{Altinoluk:2019fui}.
Consequently, the offshellness of the initial-state gluon enters both into the
TMD gluon density and to the hard matrix element.
This way, ITMD generalizes the small-$x$ TMD factorization~\cite{Dominguez:2011wm}, extending its range of
applicability from the case $Q_s\ll k_T \lesssim p_{T1},p_{T2}$ to 
$Q_s\ll p_{T1},p_{T2}$ and $Q_s < k_T < p_{T1},p_{T2}$. 
More precisely, in this work,  we employ the ITMD* version of the formalism, with the asterisk
indicating that we do not account for linearly polarized gluons in the unpolarized
target \cite{Mulders:2000sh}, whose contribution, as discussed below, is
suppressed for our choice of kinematical cuts. (See
Refs.~\cite{Metz:2011wb,Dominguez:2011br,Boer:2016fqd,Marquet:2017xwy,Altinoluk:2021ygv}
for a discussion of linearly polarized gluons in the saturation formalism and
beyond.)
The ITMD* formula for our process of interest reads
\begin{equation}
  d\sigma_{eh \to e' + 2j + X} =
  \int \frac{d x}{x}\frac{d^2 k_T}{\pi}\, \calF^{(3)}_{gg} (x, k_T, \mu)\,
  \frac{1}{4x\Pelec\!\cdot\!\Phadr}\,
  d\Phi(\Pelec,k; p_e,p_1,p_2)\,
  |\overline{M}_{eg^* \to e'+2j}|^2,
  \label{eq:ITMD}
\end{equation}
where the momenta are denoted such that $\Phadr^\mu$ refers to the initial-state
hadron, $\Pelec^\mu$ to initial-state electron, $k^\mu=x\Phadr^\mu+k_T^\mu$ to
the initial-state space-like gluon, $p_e^\mu$ to the final-state electron, and
$p_{1,2}^\mu$ to the final-state partons, see Fig.~\ref{fig:diag}.
The differential phase space element is given by
%
\begin{equation}
d\Phi(\Pelec,k; p_{e'},p_1,p_2)
=
(2\pi)^{-2}\,d^4p_e\delta_+(p_e^2)\,d^4p_1\delta_+(p_1^2)\,d^4p_2\delta_+(p_2^2)\,\delta^4(\Pelec+k-p_{e'}-p_1-p_2)
~,
\end{equation}
%
so the electrons are assumed to be massless.
In the above, the momentum of the off-shell gluon is given by
$k^{\mu}=x\Phadr^{\mu}+k_T^{\mu}$, with longitudinal momentum fraction $x$ of
nucleus $\Phadr^{\mu}$ and transverse momentum such that $\Phadr\cdot k_T=0$
. 
$|\overline{M}|^2$ is the square of the parton-level scattering amplitude summed
over spins and colors of the final-state electron and partons and averaged over the initial-state
electron spins and gluon colors. Furthermore, it is  summed over
the allowed flavors of final-state partons.
$\calF^{(3)}_{gg} (x, k_T, \mu)$ is the hard-scale-dependent
Weizs\"acker-Williams~(WW) 
unpolarized 
gluon density, counting the number of gluons at
resolution scale $\mu$. 
It is given by
\begin{equation}
    {\cal F}_{gg}^{(3)}(x,k_T,\mu)  =
    \int \!\! d b_T dk^{\prime}_T \, b_T\, k^\prime_T\, 
    J_0(b_T\,k^\prime_T)\, 
    J_0(b_T \,k_T)\,
    {\cal F}_{gg}^{(3)}(x,k^\prime_T)\, 
    e^{-S_\text{Sud}^{g\to q\bar q}(\mu,b_T)}\,,
    \label{eq:Fgg3sud}
\end{equation}
where 
\begin{equation}
\calF^{(3)}_{gg} (x,
k_T)=2\int\frac{d\xi^-d^2\xi_T}{(2\pi)^3\Phadr^+}e^{ix\Phadr^++\xi^-ik_T\cdot\xi_T}\langle
\Phadr|F^{+i}(\xi^-,\xi_T)U^{[+]\dagger} F^{+i}(0,0)U^{[+]}|\Phadr\rangle\,,
\end{equation}
is the WW gluon distribution defined in terms of the matrix element of gluon field strength tensor components  $F^{+i}$, displaced in the light cone and the transverse directions \cite{Dominguez:2010xd}. 
 $U^{[+]}$ are future-pointing gauge links, which make the gluon TMD
gauge-invariant.
The hard scale dependence of (\ref{eq:Fgg3sud}) comes from the Sudakov form
factor, which itself is a process-dependent object. For dijet production in DIS, the
Sudakov at the leading logarithmic approximation and with fixed coupling
takes the form~\cite{Mueller:2013wwa}
\begin{equation}
  S_\text{Sud}^{g\to q\bar q}(\mu,b_T)
  = \frac{\alpha_s N_c}{4\pi} \ln^2\frac{\mu^2 b_T^2}{4 e^{-2\gamma_E}}\,,
\end{equation}
where, in this work, we take $\alpha_s = 0.2$.

By using the above gluon density 
with appropriate evolution (see below) 
in the cross section formula (\ref{eq:ITMD})
we achieve simultaneous resummation of small-$x$ and the Sudakov
logarithms.  The former are relevant because by 
imposing the appropriate cuts, \ie\ selecting the forward jets, we focus the probed longitudinal fractions $x$ on its relatively small values.
The latter appear because the production of relatively hard jets introduces
additional logarithmic enhancements of the form $\ln p_T^2/k_T^2$, which
should be resummed.
As mentioned above, the factorization formula (\ref{eq:ITMD}) does not account for the linearly polarized
gluons in the unpolarized hadrons. Since we are interested exclusively in the unpolarized WW gluon distributions and the related Sudakov resummation, we shall impose a rather low $Q^2$ cut, so that $Q^2/P_T^2$ remains small and thus suppresses the linearly polarized gluon contributions.
Even though it is possible to retrieve the linearly polarized part from the CGC formulation (see \cite{Altinoluk:2021ygv,BoussarieEIC2021}), it is important to be able to test separate components of the calculation. 

The WW gluon density can be obtained as a direct solution of the  Balitsky-Jalilian-Marian-Iancu-McLerran-Weigert-Leonidov-Kovner (B-JIMWLK) evolution
equation \cite{Balitsky:1995ub,JalilianMarian:1997jx,JalilianMarian:1997gr,JalilianMarian:1997dw,Kovner:2000pt,Kovner:1999bj,Weigert:2000gi,Iancu:2000hn,Ferreiro:2001qy} or, assuming a Gaussian approximation, it can be constructed from the
dipole gluon density obeying the Balitsky-Kovchegov (BK) equation
\cite{Balitsky:1995ub,Kovchegov:1999yj}. In this paper we choose the latter
option since the higher order corrections are better understood for the BK equation and are expected to be more relevant than the simplification due to the Gaussian approximation. Using that, we compute the gluon density from the solution of the BK equation with
higher order corrections according to prescription of Kwieci\'nski, Martin and
Sta\'sto~(KMS)~\cite{Kwiecinski:1997ee,Kutak:2003bd} yielding the Kutak-Sapeta (KS)
gluon density \cite{Kutak:2012rf} fitted to the proton's $F_2$ structure function
data.
The momentum-space formulation of the BK equation allows us to treat the
kinematics exactly. 
Therefore, the $x$ variable appearing in the formulae above is not the Bjorken
$x_{\mathrm{Bj}}=Q^2/2P_h\cdot q$, but the actual fraction of the hadron momentum
carried by the scattering gluon, and these two can differ significantly.
Within the Gaussian
approximation, one derives the following formula for the WW gluon density \cite{vanHameren:2016ftb}
\begin{equation}
  \calF^{(3)}_{gg} (x,
  k_T)=\frac{2\pi^2\alpha_s}{N_ck_T^2S_\perp}\frac{1}{2}\int_{k_T^2}{dk_T^{\prime 2}}\ln\frac{k_T^{\prime
  2}}{k_T^2}
  \int\frac{d^2q_T}{q_T^2}{\cal F}_{qg}^{(1)}(x,q_T){\cal
  F}_{qg}^{(1)}(x,k_T'-q_T)\,,
  \label{eq:WWfromDip}
\end{equation}
where ${\cal F}_{qg}^{(1)}$ is the dipole gluon density and $S_\perp$ is the
target's transverse area. 

\begin{figure}[t]
  \begin{center}
    \includegraphics[width=0.8\textwidth]{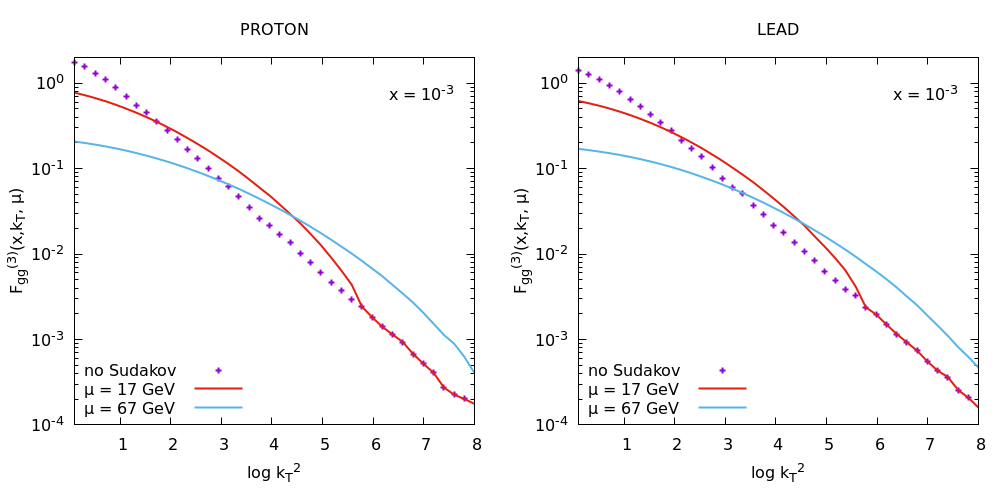}
  \end{center}
  \caption{
  The WW gluon density (\ref{eq:Fgg3sud}), in the proton (left) and lead
  (right),  with and without Sudakov resummation, as a function of the
  transverse momentum of the gluon, for various values of the hard scale.
  }
  \label{fig:Fgg3}
\end{figure}

In Fig.~\ref{fig:Fgg3} we show the WW KS gluon distributions in proton~(left)
and lead~(right), with and without Sudakov form factors, as functions of the
transverse momentum $k_T$ and the hard scale $\mu$, for one particular $x =
10^{-3}$. (The gluon density is available from the TMDlib
\cite{Abdulov:2021ivr}.) First of all, let us notice that the WW gluon distribution has no
maximum, contrary to the dipole gluon~\cite{Kutak:2012rf, vanHameren:2016ftb}.
Secondly, we see that the Sudakov factor suppresses the gluon distribution at low
$k_T$ and enhances it at higher $k_T$. Because the Sudakov form factor is derived in
the regime $\mu \propto p_T \gg k_T$, we apply it only to that part of the gluon
density where $\mu>k_T$. In the remaining domain, we use the gluon without
Sudakov, given in Eq.~(\ref{eq:WWfromDip}). This is visible in
Fig.~\ref{fig:Fgg3} as a kink of the curve corresponding to $\mu = 17\, \GeV$.
(A similar kink exits also for the $\mu = 67\, \GeV$ curve but it is located at
larger values of $\log k_T^2$.)

\section{Results}

\begin{figure}[t]
  \begin{center}
   \includegraphics[width=0.49\textwidth]{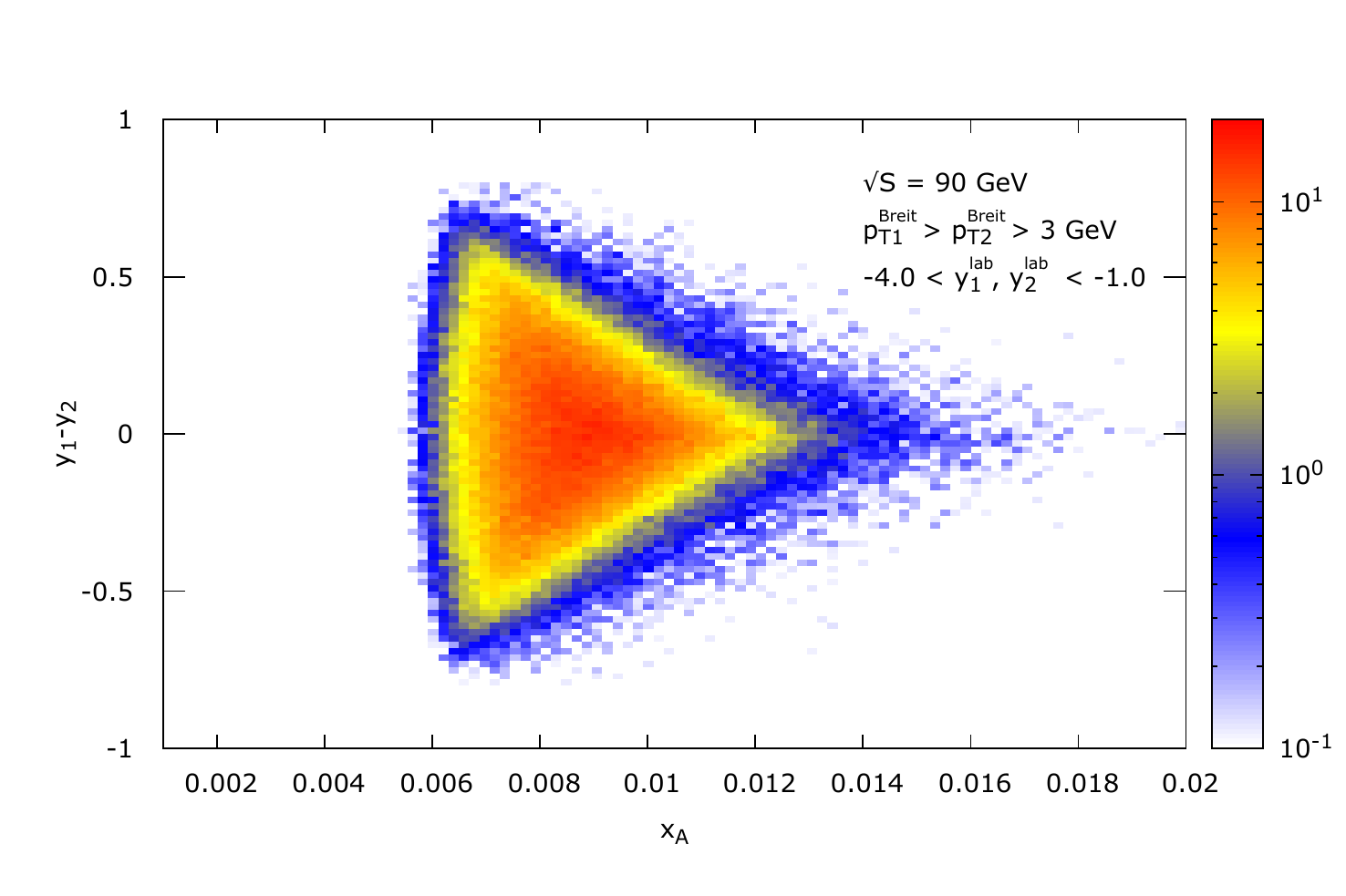}
   \includegraphics[width=0.49\textwidth]{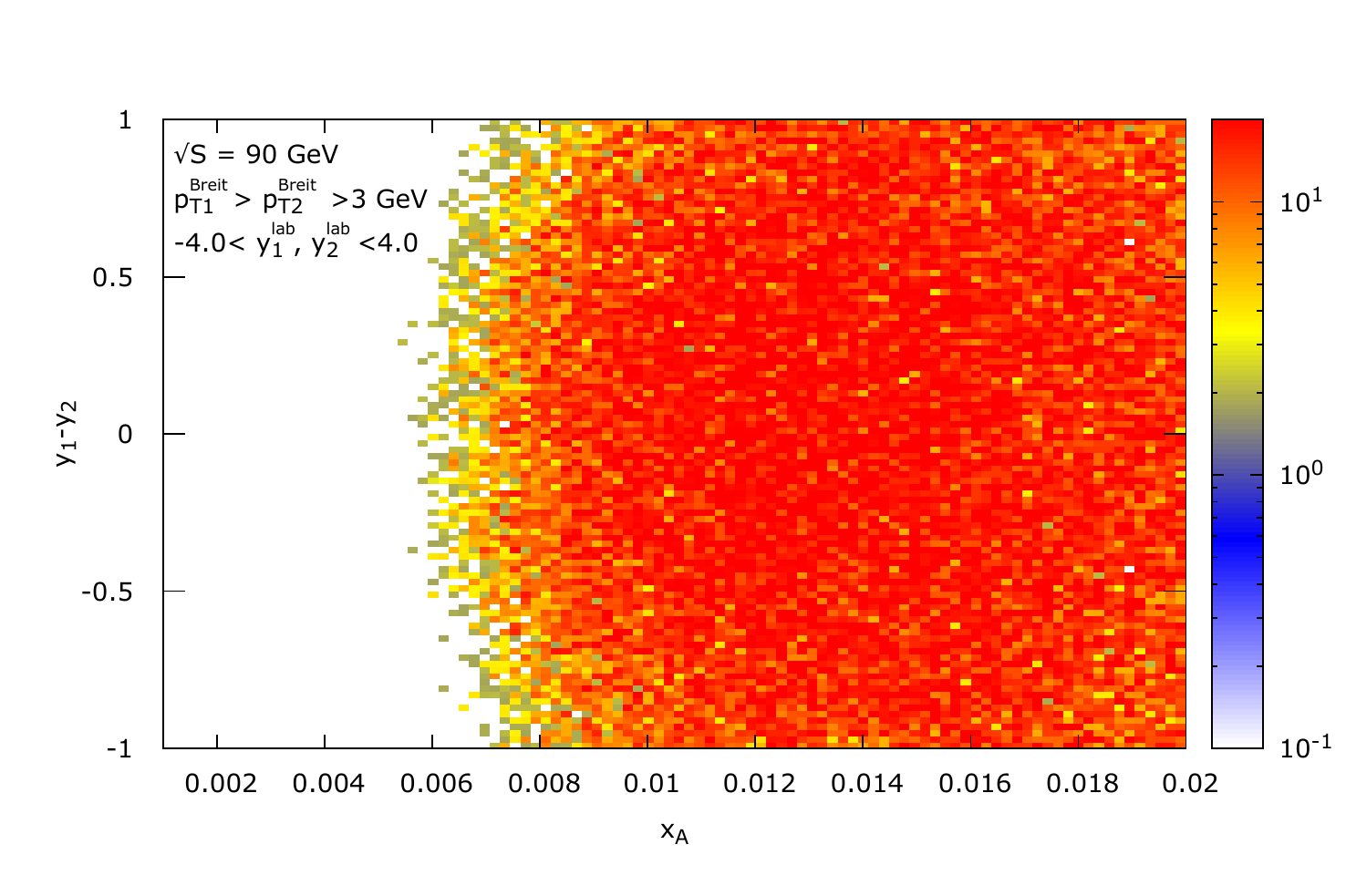}
  \end{center}
  \caption{ Density plots representing contribution of events for a given gluon $x$ and jet rapidity difference for the asymmetric rapidity cuts (left) and the symmetric rapidity cuts (right). The asymmetric rapidity cuts  guarantee good focusing of the cross section around smaller values of $x$ required by the formalism.
  }
  \label{fig:rapiditycuts}
\end{figure}

\begin{figure}[t]
  \begin{center}
   \includegraphics[width=0.4\textwidth]{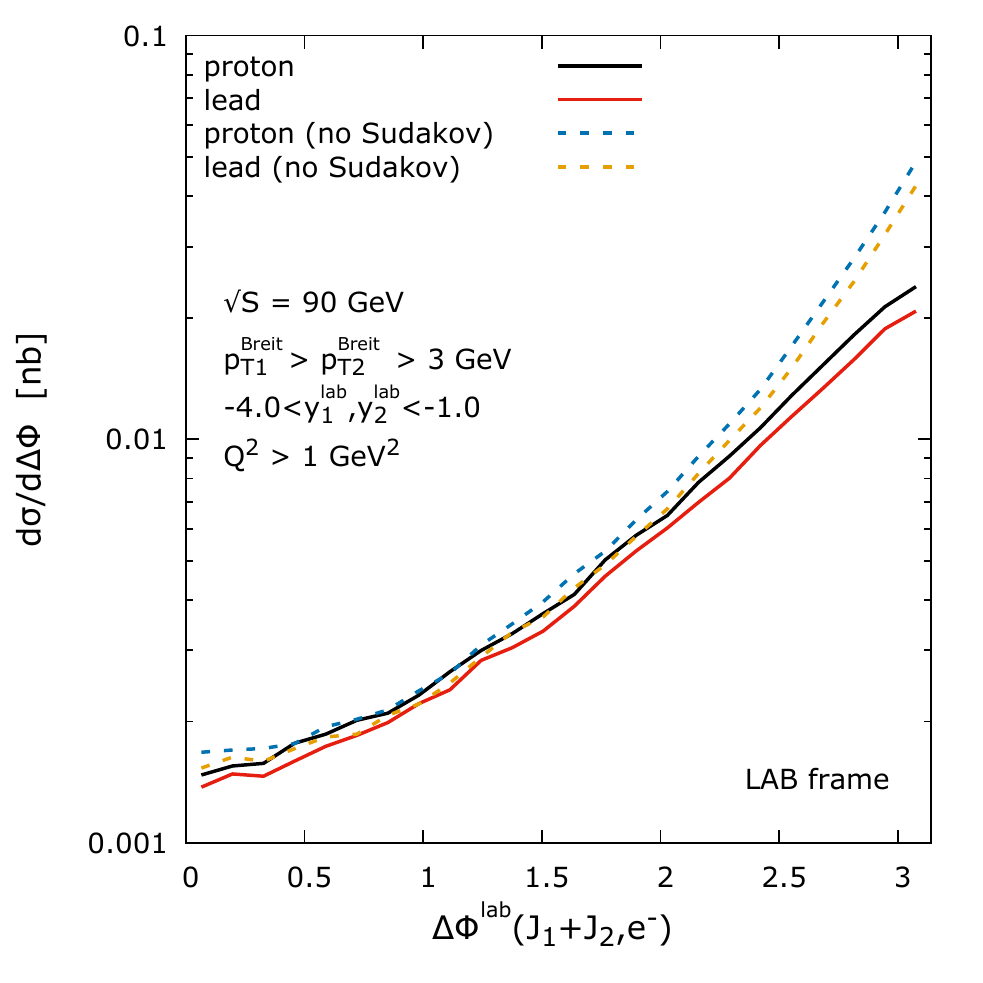}
   \includegraphics[width=0.4\textwidth]{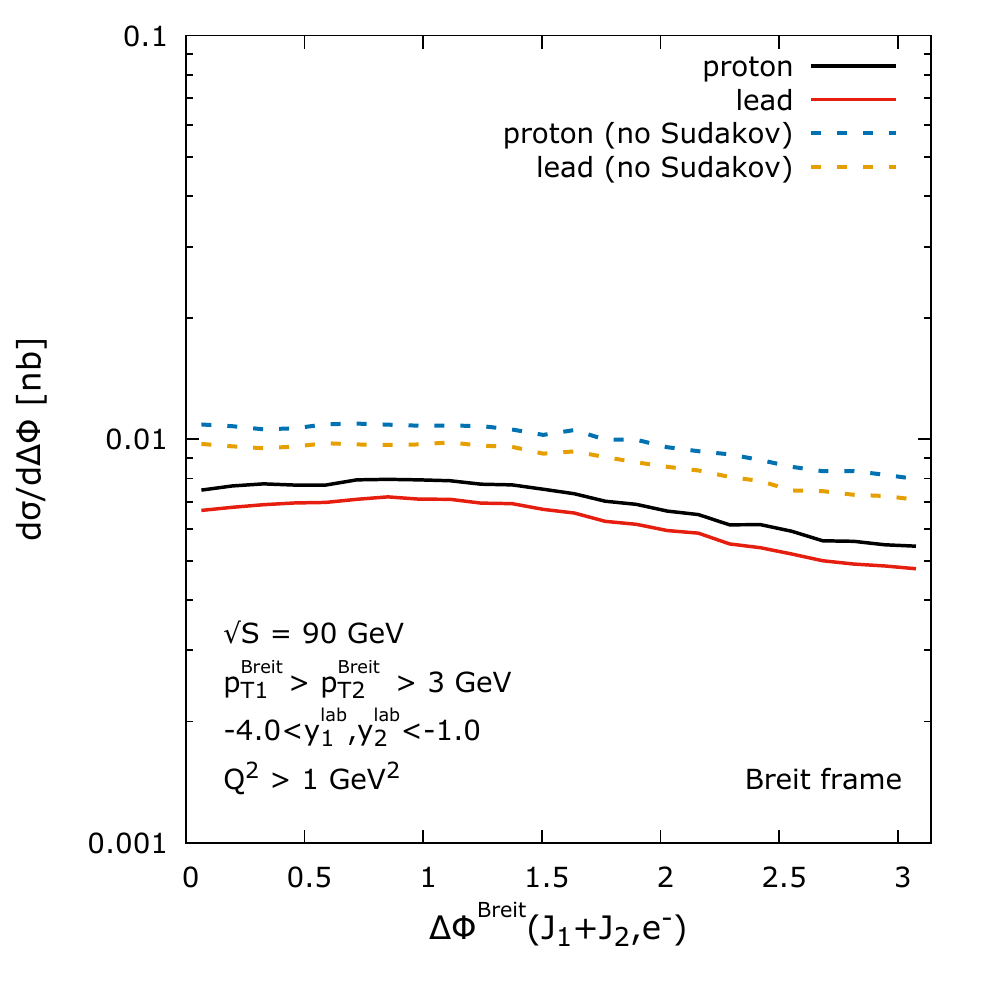}
  \end{center}
  \caption{Azimuthal correlations between the total transverse momentum of the
  dijet system and the transverse momentum of the scattered electron at EIC in
  two frames: the LAB frame (left), the Breit frame (right). The calculation has
  been done within the ITMD* framework with the Weizs\"acker-Williams gluon distribution obtained from the Kutak-Sapeta fit to HERA data.
  }
  \label{fig:angleplot}
\end{figure}

In this section we present numerical results for the differential cross section
as a function of the azimuthal angle between the total transverse momentum of
the dijet system and the transverse momentum of the scattered electron, see the
right plot of Fig.~\ref{fig:diag}.  The calculations are performed both for the $e-p$
and $e-Pb$ collisions at the center of mass (CM) energy $\sqrt{S}=90\, \GeV$ per
nucleon. We look at low-virtuality events, with $Q^2>1\,  \GeV^2$, with
inelasticity $0.1<\nu<0.85$. The final-state partons are subject to a jet
algorithm, which, at leading order, is simply the cut on the azimuthal
angle-rapidity plane, \ie\ the requirement that $\sqrt{\Delta\Phi^2 + \Delta
y^2}<R$. We choose the jet radius $R=1$, as suggested \eg\ in
Ref.~\cite{Arratia:2019vju}. The jet definition is imposed in the Breit frame.
Furthermore, we require that the transverse momenta of the jets satisfy
$p_{T1}>p_{T2}>3\, \GeV$ in the Breit frame.  In order to have a quite narrow
distribution of the longitudinal fraction $x$, we select events with the
laboratory (LAB) frame rapidities $-4 < y_1,y_2 < -1$,
where the negative rapidity correspond to the incoming electron beam. 
The reason for this is illustrated
in Fig.~\ref{fig:rapiditycuts}, where we show density plots for the gluon's $x$
versus the rapidity difference $y_1-y_2$, for the above asymmetric cut (left)
versus the symmetric cut $-4<y_1,y_2<+4$ (right). We see that the asymmetric cut
gives us a very good focusing of the cross section around smaller values of the
gluon $x\sim  0.008$. Note, that this corresponds to quite low Bjorken
$x_{\mathrm{Bj}}\lesssim 10^{-4}$. On the contrary, the symmetric cut provides a
rather broad distribution in the gluon's~$x$, extending towards large values, where
the formalism is questionable.  To summarize, we impose the following cuts:
\begin{eqnarray}
  & \sqrt{S}=90\, \GeV,\qquad Q^2 > 1\, \GeV^2, \qquad 0.1<\nu<0.85 \notag\\ 
  & \Delta R^{\mathrm{Breit}}<1, \qquad p^{\mathrm{Breit}}_{T1}>p^{\mathrm{Breit}}_{T2}>3\, \GeV, 
  \label{eq:sel}
\\
  & \quad -4<y^{\mathrm{lab}}_1,y^{\mathrm{lab}}_2<-1. \notag
\end{eqnarray}
We generate events using the {\tt KaTie} \cite{vanHameren:2016kkz} Monte Carlo
and present our results both in the laboratory~(LAB) frame and the Breit frame.
The second option is a
preferred choice for dijet observables since it suppresses the LO contributions,
\ie\ processes dominated by a quark jet~\cite{Liu:2018trl}.  

\begin{wrapfigure}{r}{0.38\textwidth}
  \vspace{-20pt}
  \begin{center}
    \includegraphics[width=0.38\textwidth]{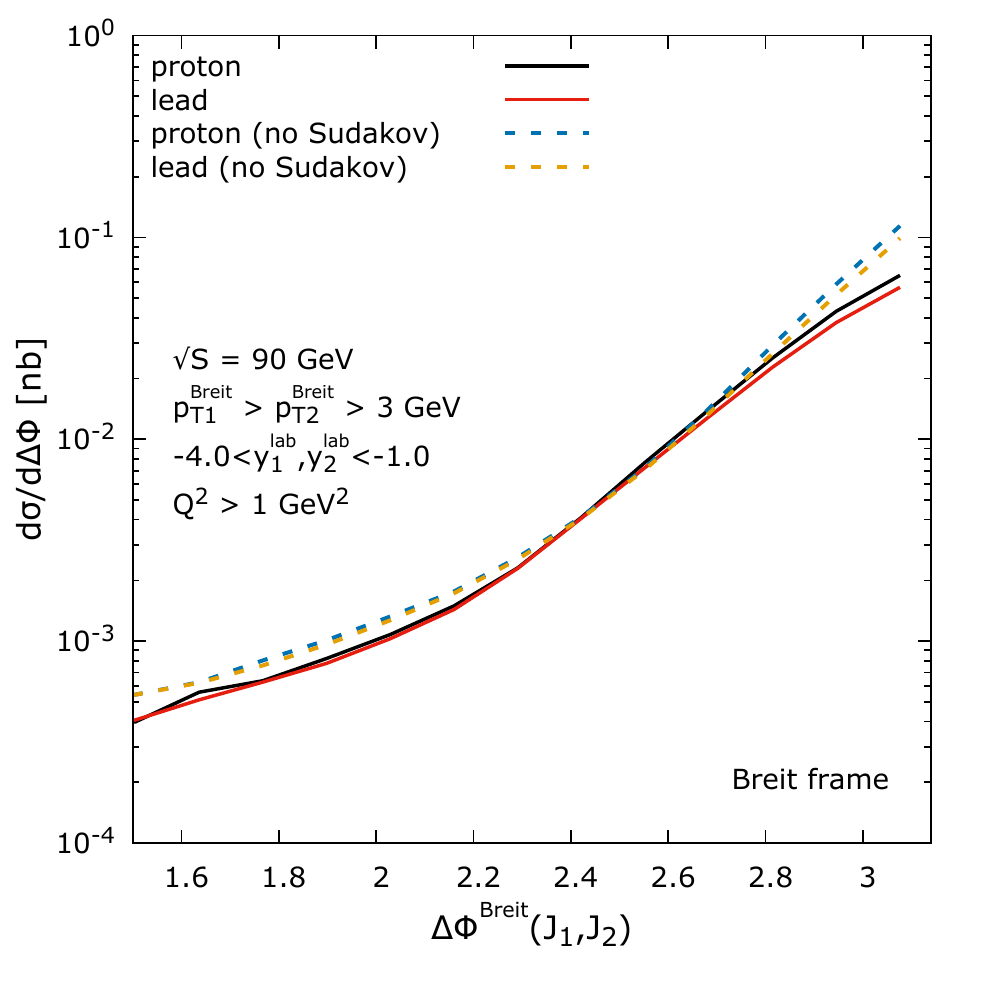}
  \end{center}
  \caption{Azimuthal correlations between the jets in the Breit frame. 
  \label{fig:dijetcor}
  }
\end{wrapfigure}

As explained in the previous section, the selection choice~(\ref{eq:sel}), with $p_{T}$'s of jets larger than the $Q_s$,  is motivated by
our goal to study the Sudakov resummation effects together with the kinematical
twist \cite{Altinoluk:2019fui}, in addition to the saturation effects, which, although rather mild, are expected to be present.
In other words, the genuine twist configurations (\ie\ 
multiple exchanges of non-soft gluons between the remnant and dijets~\cite{Altinoluk:2019wyu}) as
well as contributions from linearly polarized gluons \cite{Marquet:2017xwy} are
suppressed. 
For the latter, we do not impose an upper cut on $Q^2$ as the cross section is strongly peaked at low virtualities and such additional cut would introduce only a small correction.

Our predictions for the cross section as a function of the angle between
electron and dijet system are presented in Fig.~\ref{fig:angleplot}. We plot also a
control result based on a calculation that neglects the Sudakov form factor. The
comparison of the two results clearly shows that while saturation effects are
mild, the Sudakov effects are fairly large. 
This feature is clearly visible in both the LAB and the Breit frame. In the LAB
frame the difference is concentrated around the correlation region of the dijet-electron system. 
In the Breit frame, on the other hand, the Sudakov form factor suppresses the cross section over the whole region of the azimuthal angle. Let us note, that the Breit frame is defined in a standard way in our calculation, that is by requiring the photon to have nonzero only the spatial $z$ component. However, unlike in collinear factorization, there is a non-zero transverse momentum of the incoming parton in the factorization formula (\ref{eq:ITMD}), which makes this frame slightly less intuitive.
The saturation effects are observed as an almost constant suppression of the electron-lead cross section (normalized to the number of nucleons) over all azimuthal angles. The suppression reaches maximally about 15\%, which although modest, is an observable effect. 
To be so, the jets must have relatively small $p_T\sim 3\, \GeV$ in Breit frame,
thus, in practice, one can consider dihadron-electron correlations.

\section{Conclusions}

In this Letter we proposed a new study of the Weizs\"acker-Williams
gluon distribution, using the azimuthal correlations between the forward dijet system 
and the scattered electron at the EIC. We chose the kinematic cuts such that
both the genuine twists and the linearly polarized gluons are suppressed. This gives
us a direct access to the unpolarized distribution via the small-$x$ ITMD factorization. 
We provided predictions over whole range of the azimuthal angle, which is possible thanks to the inclusion 
of the kinematic twists and full phase space. 

We found that forward dijet-electron azimuthal correlations provide a more
sensitive observable to the Sudakov suppression, shown in
Fig.~\ref{fig:angleplot}, as compared  
to the jet-jet correlations alone, depicted in Fig~.\ref{fig:dijetcor}. 
In the latter case the difference is concentrated in the
region very close to the correlation peak (in 
Breit frame), whereas for the former
observable it extends over large region of azimuthal angle.  

Our results may serve as a guidance for future measurements at EIC and 
improved calculations at the NLO accuracy.

\section*{Acknowledgement}
We acknowledge Hannes Jung for useful correspondence.
KK acknowledges informative email exchange with Miguel Arratia.
PK is thankful to Farid Salazar for discussions.
AvH is supported by grant no.\ 2019/35/B/ST2/03531 of the Polish National Science Centre.
PK is partially supported by National Science Center, Poland, grant no. 2018/31/D/ST2/02731.
KK is partially supported by grant no.\ DEC-2017/27/B/ST2/01985. 
SS is partially supported by the Polish National Science Centre grant no.
2017/27/B/ST2/02004.
\bibliographystyle{apsrev4-2}
\bibliography{references}

\end{document}